\documentclass[preprint,12pt]{elsarticle}
\usepackage{amssymb}
\journal{High Energy Density Physics}
\begin{document}
\begin{frontmatter}
\title{Ion-acoustic solitary pulses in a dense plasma}
\author{B. Hosen$^{1*}$, M. G. Shah$^{2}$, M. R. Hossen$^{3}$, and A. A. Mamun$^{1}$}

\address{$^{1}$Department of Physics, Jahangirnagar University,
Savar, Dhaka-1342, Bangladesh.\\$^{2}$Department of Physics, Hajee
Mohammad Danesh Science and Technology University, Dinajpur-5200,
Bangladesh.\\$^{3}$Department of General Educational Development,
Daffodil International University, Dhanmondi, Dhaka-1207,
Bangladesh.} \ead{$^*$hosen.plasma@gmail.com}

\begin{abstract}
The propagation of ion-acoustic solitary waves (IASWs) in a
magnetized, collisionless degenerate plasma system for describing
collective plasma oscillations in dense quantum plasmas with
relativistically degenerate electrons, oppositely charged
inertial ions, and positively charged immobile heavy elements is
investigated theoretically. The perturbations of the magnetized
quantum plasma are studied employing the reductive perturbation
technique to derive the Korteweg-de Vries (K-dV) and the modified
K-dV (mK-dV) equations that admits solitary wave solutions.
Chandrasekhar limits are used to investigate the degeneracy
effects of interstellar compact objects through equation of state
for degenerate electrons in case of non-relativistic and
ultra-relativistic cases. The basic properties of small but
finite-amplitude IASWs are modified significantly by the combined
effects of the degenerate electron number density, pair ion number
density, static heavy element number density and magnetic field.
It is found that the obliqueness affects both the amplitude and
width of the solitary waves, whereas the other parameters mainly
influence the width of the solitons. The results presented in
this paper can be useful for future investigations of
astrophysical multi-ion plasmas.
\end{abstract}


\end{frontmatter}

\section{Introduction}

The study of nonlinear wave phenomena in pair ion plasmas is
predominantly significant due to its application in space and
laboratory plasmas
\cite{Tiwari1980,D'Angelo1990,Chen2001,Nakamura2003,Mamun2009,Perrone2013,Shahmansouri2015}.
Generally, pair ion plasma system is a system containing more
than one types of ions and has a great importance to different
field of plasma science and technology. The evolution of solitons
in plasma having both positive and negative ion species have been
investigated by Rizzato \textit{et al.} \cite{Rizzato1987}. In
different laboratory situations, (viz. plasma processing
reactors, neutral beam sources, low-temperature laboratory
experiments, etc.) the existence of positive-negative ion plasmas
has also been found \cite{Gottscho1986,Bacal1979,Jacquinot1977}.
In addition, a new experimental setup is developed for ion energy
loss measurements in a partially ionized, moderately coupled
carbon dense plasma with the presence of heavy ion beam
\cite{Ortner2016}. Therefore, the study of plasma excitations in
magnetized dense plasmas allow us learning different basic wave
phenomena, such as solitons, shock waves, double layers,
vortices, etc. One of the basic wave processes namely
ion-acoustic (IA) waves in celestial as well as terrestrial
plasma system have been studied for several decades both
theoretically and experimentally
\cite{Washimi1966,Sagdeev1966,Ikezi1970,Mase1975,Baboolal1990,Bharuthram1992,Popel1995,Mamun1997,Mamun2001}.
Washimi and Taniuti \cite{Washimi1966} were the first to derive
the Korteweg de-Vries (KdV) equation governing the propagation of
IA waves in a collisionless plasma. IA waves have been studied in
various cases, i.e., multi-ion plasma compositions
\cite{Bharuthram1992,Shahmansouri2014}, two temperature electron
plasmas
\cite{Baboolal1990,Shahmansouri2014,Baboolal1989,Rice1993,Ghosh1996},
plasma with superthermal electrons \cite{Shah2015d,Shah2016},
pair-ion plasmas \cite{Popel1995,Alinejad2011}, degenerate
plasmas \cite{Hossen2014} and relativistic effect of plasmas
\cite{Shah2015c}. It is well-known that the external magnetic
field can modify the propagation properties of the electrostatic
IA solitary structures. The effect of an ambient external
magnetic field on the electrostatic waves has been studied by a
number of authors \cite{Yu1980,Shukla2001,Mamun1998,Sultana2012}.
Yu \textit{et al.} \cite{Yu1980} extended the Sagdeev approach to
study the IA waves in a magnetized plasma. Their results show how
the external magnetic field affects the nature of the solitary
wave profiles.

Currently, theoretical concerns are to analyze the environment of
the compact objects, such as white dwarfs, neutron stars, etc.
\cite{Chandrasekhar1931,Chandrasekhar1935,Shapiro1983,Shah2015a,Shah2015b}.
The basic constituents of white dwarfs are mainly oxygen, carbon,
helium with an envelope of hydrogen gas. The icy satellites of
Saturn have been shown to be the source of the heavy positive ion
i.e., O$^+$, N$^+$, etc. plasma in the inner Saturnian torus
\cite{Eviatar 1983,Richardson 1986,Ema2015}. The degenerate
electron number density in such a compact object is so high (e.g.
the degenerate electron number density can be of the order of
$10^{30}$ $cm^{-3}$ in white dwarfs, and of the order of
$10^{36}$ $cm^{-3}$ in neutron stars)
\cite{Hossen2014,Hossen2014a} that the electron Fermi energy is
comparable to the electron mass energy and the electron speed is
comparable to the speed of light in a vacuum. Within
astrophysical objects, the lower energy state is filled with
electrons so additional electrons cannot give up energy to the
lower energy state and they generate degeneracy pressure which is
explained by the joined effects of Pauli's exclusion principle and
Heisenberg's uncertainty principle. The equation of state for
degenerate electrons in such interstellar compact objects are
explained by Chandrasekhar for two limits where
$({P_e}\propto{n_e}^{5/3})$ for the non-relativistic limit and
$({P_e}\propto{n_e}^{4/3})$ for the ultra-relativistic limit,
where $P_e$ is the degenerate electron pressure and $n_e$ is the
degenerate electron number density. To demonstrate the equation
of state, Chandrasekhar introduced that for the nonrelativistic
degenerate electrons,
$\gamma=\frac{5}{3};~K=\frac{3}{5}\left(\frac{\pi}{3}\right)^{\frac{1}{3}}
\frac{\pi\hbar^2}{m}\simeq\frac{3}{5}\Lambda_c\hbar c$ where $K$
is the proportionality constant, $\Lambda_c=\pi \hbar/mc=1.2\times
10^{-10}~cm$, and $\hbar$ is the Planck constant divided by
$2\pi$ and for the ultrarelativistic degenerate electrons,
$\gamma=\frac{4}{3};~K=\frac{3}{4}\left(\frac{\pi^2}{9}\right)^{\frac{1}{3}}
\hbar c\simeq\frac{3}{4}\hbar c$
\cite{Chandrasekhar1931,Chandrasekhar1935}.

Now, a large number of authors studied the basic properties of
solitons or shock waves by deriving the K-dV, mK-dV, Gardner or
Burgers equation for planar or nonplanar cases in considering
different types of effects
\cite{Mamun2009,Chatterjee2012,Mannan2012,El-Taibany2007,Zobaer2012,Samanta2013,Saha2014,Shah2015m}.
Alinejad and Mamun \cite{Alinejad2011} studied oblique
propagation of small amplitude IA solitons in a pair plasma with
superthermal electrons. The nonextensivity effects on the
obliquely propagating IA waves in a magnetized plasma have been
investigated by Shahmansouri and Alinejad
\cite{Shahmansouri2013}. The effect of an applied uniform
magnetic field on the propagation of magnetosonic solitary waves
in the weakly relativistic limit in a magnetized multi-ion plasma
composed of electrons, light ions and heavy ions has been studied
by Wang \textit{et al.} \cite{Wang2009}. Masood \textit{et al.}
\cite{Masood2010} considered a degenerate quantum magnetized
plasma to study the propagation of electromagnetic wave. Hossen
\textit{et al.} \cite{rasel2014a,rasel2014b,rasel2014c,rasel2015}
investigated the basic features of different nonlinear acoustic
waves in the presence of heavy elements in a relativistic
degenerate plasma system that is valid only for the unmagnetized
case.

To the best of our knowledge, there are no investigations, which
have been made for condition of matter by considering magnetized
degenerate electrons and oppositely charged ions in a magnetized
quantum plasma. Therefore, in this work our main intention is to
study the basic features of IASWs by deriving the magnetized K-dV
and magnetized mK-dV equations in magnetized dense plasmas.
\section{Theoretical model and basic equations}
We consider a degenerate, dense, magnetized quantum multi-ion
plasma system consisting of both non-relativistic and
ultra-relativistic degenerate electrons, non-relativistic
degenerate inertial ions of both positively and negatively
charged and positively charged immobile heavy elements. It is
noted here that the behavior of such a multi-ion plasma may
significantly differ from the behavior of a single-ion-species
plasma. The study of the effect of magnetic lines of force of both
the magnetized positively and negatively charged ions is very
common in literature
\cite{Tiwari1980,Shahmansouri2015,Wang2009,Chakrabarti2002,Tskhakaya2005,Haider2014}.
In equilibrium, we have $n_{e0} + Z_-n_{-0}=Z_+n_{+0} +
Z_hn_{h0}$, where $Z_+$ is the number of positive ions, $Z_-$ is
the number of negative ions, $Z_h$ is the number of positive ions
residing on the heavy ion's surface and $n_{+0}$, $n_{-0}$,
$n_{e0}$ and $n_{h0}$ are the number densities of positive ions,
negative ions, electrons and heavy elements in equilibrium. The
positively charged static heavy elements participate only in
maintaining the quasi-neutrality condition in equilibrium. We
consider that the number densities of positive and negative ions
is equal in equilibrium i.e., $n_{+0}=n_{-0}$. The dynamics of
nonlinear IA waves in the presence of the external magnetic field
$\textbf{B}=\hat{z}B_0$ is governed by the following momentum
equation
\begin{eqnarray}
&&\hspace*{-10mm}\nabla\phi-\frac{K_1}{n_{e}}\nabla{n_e}^\gamma=0,\label{C1a}
\end{eqnarray}
and the non-degenerate inertial ion equations composed of the ion
continuity and ion momentum equations are given by
\begin{eqnarray}
&&\frac{\partial n_+}{\partial t} + \nabla .({n_+}{\textbf{u}_+})=
0,
\label{C1b}\\
&&\frac{\partial n_-}{\partial t} + \nabla .({n_-}{\textbf{u}_-})=
0,
\label{C1c}\\
&&\frac{\partial \textbf{u}_+}{\partial t} +
({\textbf{u}_+}.\nabla)u_+=- \nabla \phi+
\omega_{c_{+}}({\textbf{u}_+}\times{\hat{z}}),\label{C1d}\\
&&\frac{\partial \textbf{u}_-}{\partial t} +
({\textbf{u}_-}.\nabla)u_-=\beta \nabla \phi+
\omega_{c_{-}}({\textbf{u}_-}\times{\hat{z}}),\label{C1e}
\end{eqnarray}
The equation that is closed by Poisson's equation
\begin{eqnarray}
&&\nabla^2 \phi=\delta{n_e}+\sigma{n_-}-{n_+}-{Z_h}\mu,\label{C1f}
\end{eqnarray}
where $n_+$, $n_-$ and $n_e$ is the perturbed number densities of
inertial positive ions, inertial negative ions and degenerate
electrons, respectively). $\textbf{u}_i$ is the plasma species
fluid speed normalized by $C_{pm}=(m_ec^2/Z_pm_p)^{1/2}$ with
$m_e$ ($m_p$) being the electron (plasma ion specie's) rest mass,
$c$ is the speed of light in vacuum, $\phi$ is the electrostatic
wave potential normalized by $m_ec^2/e$ with e being the
magnitude of the charge of an electron. Here $\beta =
(Z_-m_+/Z_+m_-)$ is the ratio of the masses of the positive and
the negative ion multiplied by their charge per ion $Z_{s}$
(where s = +, -), $\delta\ = (n_{e0}/Z_+n_{+0})$ is the ratio of
the number density of electron and positive ion multiplied by
charge per positive ion $Z_+$, $\sigma = (Z_-n_{-0}/Z_+n_{+0})$
is the ratio of number density of negative and positive ion
multiplied by their charge per ion and $\mu = (n_{h0}/Z_+n_{+0})$
is the ratio of the number density of heavy element and positive
ion multiplied by $Z_+$. The nonlinear propagation of usual IA
waves in electron-ion plasma can be recovered by setting $\mu=0$.
The time variable ($t$) is normalized by ${\omega_{pm}}=(4 \pi
n_{p0}e^2/m_p)^{1/2}$, and the space variable ($x$) is normalized
by $\lambda_{m}=(m_ec^2/4 \pi n_{p0}e^2)^{1/2}$. We have defined
the parameter that appears in Eq. (1) as
$K=n_{e0}^{\gamma-1}K_e/{m_e}{c}^2$.

\section{Derivation of the Magnetized K-dV Equation}

In order to investigate the dynamics of small but finite
amplitude obliquely propagating IA waves in the relativistic
degenerate dense magnetized multi-ion quantum plasma, we use the
standard reductive perturbation technique to drive the K-dV
equation. We now introduce the new set of stretched coordinates as

\begin{eqnarray}
&&\eta=\epsilon^{1/2}(L_xx+L_yy+L_zz-V_pt),
\label{C2a}\\
&&T={\epsilon}^{3/2}t, \label{C2b}
\end{eqnarray}

\noindent where $\epsilon$ is a smallness parameter $(0 < \epsilon
< 1)$ measuring the amplitude of perturbation, $V_p$ is the wave
phase velocity normalized by the IA speed ($C_{im}$), and $l_x$,
$l_y$, and $l_z$ are the directional cosines of the wave vector k
along the x, y, and, z axes, respectively, so that $l_x^2$ +
$l_y^2$ + $l_z^2$ = 1. It is noted here that x, y, z are all
normalized by the Debye length $\lambda_{D}$, and $T$ is
normalized by the inverse of ion plasma frequency
($\omega_{pi}^{-1}$ ). We may expand $n_s$, $u_s$, and $\phi$ in
power series of $\epsilon$ as

\begin{eqnarray}
&&n_s=1+\epsilon n_s^{(1)}+\epsilon^{2}n_s^{(2)}+ \cdot \cdot
\cdot, \label{C2c}\\
&&u_{ix,y}=0+\epsilon^{3/2}
u_{ix,y}^{(1)}+\epsilon^{2}u_{ix,y}^{(2)}+\cdot \cdot \cdot,
\label{C2d}\\
&&u_{iz}=0+\epsilon u_{iz}^{(1)}+\epsilon^{2}u_{iz}^{(2)}+\cdot
\cdot \cdot,
\label{C2e}\\
&&\phi=0+\epsilon\phi^{(1)}+\epsilon^{2}\phi^{(2)}+\cdot \cdot
\cdot, \label{C2f}
\end{eqnarray}

Now, substituting Eqs. (\ref{C2a}) - (\ref{C2f}) into Eqs.
(\ref{C1a}) - (\ref{C1f}) and taking the lowest order coefficient
of $\epsilon$, we obtain, $u_{+z}^{(1)}={L_z \phi^{(1)}}/{V_p}$,
$u_{-z}^{(1)}={-L_z \beta\phi^{(1)}}/{V_p}$,
 $n_{+}^{(1)}={L_z^2 \phi^{(1)}}/{V_p^2}$, $n_{-}^{(1)}={-L_z^2 \beta\phi^{(1)}}/{V_p^2}$,
 $n_{e}^{(1)}=\phi^{(1)}/K_{11}$, and
$V_p=L_z\sqrt{{\frac{K_{11}(\sigma\beta+1)}{\delta}}}$ represents
the dispersion relation for the IA waves that move along the
propagation vector $k$.

To the lowest order of x- and y-component of the momentum
equations (\ref{C1d}) and (\ref{C1e}) we get,

\begin{eqnarray}
&&u_{+y}^{(1)}=\frac{L_x}{\omega_{c+}}\frac{\partial\phi^{(1)}}{\partial\eta},
\label{C3a}\\
 &&u_{+x}^{(1)}=-\frac{L_y}{\omega_{c+}}\frac{\partial\phi^{(1)}}{\partial\eta},
 \label{C3b}\\
 &&u_{-y}^{(1)}=-\beta\frac{L_x}{\omega_{c-}}\frac{\partial\phi^{(1)}}{\partial\eta},
\label{C3c}\\
 &&u_{-x}^{(1)}=\beta\frac{L_y}{\omega_{c-}}\frac{\partial\phi^{(1)}}{\partial\eta},
 \label{C3b}
\end{eqnarray}

Now, substituting Eqs. (\ref{C2a})-(\ref{C3b}) into (\ref{C1d})
and (\ref{C1e}) one can obtain from the higher order series of
$\epsilon$ of the momentum and Poisson's equations as

\begin{eqnarray}
 &&u_{+y}^{(2)}=\frac{L_yV_P}{\omega_{c+}^2}\frac{\partial^2\phi^{(1)}}{\partial\eta^2},
 \label{C4a}\\
 &&u_{+x}^{(2)}=\frac{L_xV_P}{\omega_{c+}^2}\frac{\partial^2\phi^{(1)}}{\partial\eta^2},
\label{C4b}\\
 &&u_{-y}^{(2)}=-\beta\frac{L_yV_P}{\omega_{c-}^2}\frac{\partial^2\phi^{(1)}}{\partial\eta^2},
 \label{C4c}\\
 &&u_{-x}^{(2)}=-\beta\frac{L_xV_P}{\omega_{c-}^2}\frac{\partial^2\phi^{(1)}}{\partial\eta^2},
\label{C4d}\\
&&\frac{\partial^2\phi^{(1)}}{\partial \eta^2}=\delta
n_e^{(2)}+\sigma n_-^{(2)}-n_+^{(2)}, \label{C4e}
\end{eqnarray}

Using the same process, we get the next higher order continuity
equation as well as the z-component of the momentum equation. Now,
combining these higher order equations together with Eqs.
(\ref{C3a})-(\ref{C4e}) one can obtain

\begin{eqnarray}
&&\frac{\partial\phi^{(1)}}{\partial T} + \lambda \phi^{(1)}
\frac{\partial \phi^{(1)}}{\partial \eta}+ \beta \frac{\partial^3
\phi^{(1)}}{\partial \eta^3}=0, \label{C5a}
\end{eqnarray}

This is well-known K-dV equation that describes the obliquely
propagating IA waves in a magnetized quantum plasma.

where
\begin{eqnarray}
&&\lambda=\frac{K_{11}V_p}{2\delta}\left[\frac{{\delta}(\gamma-2)}{K_{11}^2}-\frac{{3\sigma\beta^2L_z^4}}{V_p^4}+\frac{3L_z^4}{V_p^4}\right]
\label{C5b}\\
&&\beta=\frac{K_{11}V_p}{2\delta}\left[1+\frac{\sigma\beta(1-L_z^2)}{\omega_{c-}^2}+\frac{(1-L_z^2)}{\omega_{c+}^2}\right]
\label{C5c}
\end{eqnarray}

In order to indicate the influence of different plasma parameters
on the propagation of solitary waves in magnetized quantum
plasma, we derive the solution of K-dV equation (\ref{C5a}). The
stationary solitary wave solution of standard K-dV equation is
obtained by considering a frame $\xi=\eta-u_{0}T$ (moving with
speed $u_{0}$) and the solution is,
\begin{eqnarray}
{\rm \phi^{(1)}}=\rm \phi_m{\rm[sech^{2}}(\frac{\xi}{\Delta})],
\label{solK-dV}
\end{eqnarray}

\noindent where the amplitude, $\phi_m=3u_{0}/\lambda$, and the
width, $\Delta=(4\beta/u_{0})^{1/2}$

\section{Derivation of the Magnetized mK-dV Equation}
To obtain the mK-dV equation, the same stretched co-ordinates are
applied as we used in K-dV equation in section 3 (i.e.,
Eqs.(\ref{C2a}) and (\ref{C2b})) and also used the dependent
variables which are expanded as

\begin{eqnarray}
&&\hspace*{-10mm}n_s=1+\epsilon^{1/2} n_s^{(1)}+\epsilon
n_s^{(2)}+\epsilon^{3/2} n_s^{(3)} +\cdot \cdot
\cdot, \label{C6a}\\
&&\hspace*{-10mm}u_{ix,y}=0+\epsilon
u_{ix,y}^{(1)}+\epsilon^{3/2}u_{ix,y}^{(2)}+\epsilon^{2}u_{ix,y}^{(3)}+\cdot
\cdot \cdot,
\label{C6b}\\
&&\hspace*{-10mm}u_{iz}=0+\epsilon^{1/2} u_{iz}^{(1)}+\epsilon
u_{iz}^{(2)}+\epsilon^{3/2} u_{iz}^{(3)}+\cdot \cdot \cdot,
\label{C6c}\\
&&\hspace*{-10mm}\phi=0+\epsilon^{1/2}\phi^{(1)}+\epsilon\phi^{(2)}+\epsilon^{3/2}\phi^{(3)}+\cdot
\cdot \cdot, \label{C6d}
\end{eqnarray}
We find the same expressions for $n_+^{(1)}$, $n_-^{(1)}$,
$n_{e}^{(1)}$, $u_{+z}^{(1)}$, $u_{-z}^{(1)}$, $u_{+x,y}^{(1)}$,
$u_{-x,y}^{(1)}$, $u_{+x,y}^{(2)}$, $u_{-x,y}^{(2)}$ and $V_p$ by
using the values of $\eta$ and $T$ in Eqs.(\ref{C1a})-(\ref{C1f})
and (\ref{C6a})-(\ref{C6d}) as before in section III. The next
higher order series of $\epsilon$ of continuity, momentum and
poisson's equations as

\begin{eqnarray}
&&\hspace*{-6mm}\frac{\partial n_+^{(1)}}{\partial
T}-V_p\frac{\partial n_+^{(3)}}{\partial\eta}+L_x\frac{\partial
u_{+x}^{(2)}}{\partial\eta}+L_x\frac{\partial}{\partial\eta}(n_+^{(1)}u_{+x}^{(1)})+L_y\frac{\partial
u_{+y}^{(2)}}{\partial\eta}\nonumber\\
&&\hspace*{-6mm}+L_y\frac{\partial}{\partial\eta}(n_+^{(1)}u_{+y}^{(1)})+L_z\frac{\partial
u_{+z}^{(3)}}{\partial\eta}
+L_z\frac{\partial}{\partial\eta}(n_+^{(1)}u_{+z}^{(2)})+L_z\frac{\partial}{\partial\eta}(n_+^{(2)}u_{+z}^{(1)})=0,
\label{C7a}\\
&&\hspace*{-6mm}\frac{\partial n_-^{(1)}}{\partial
T}-V_p\frac{\partial n_-^{(3)}}{\partial\eta}+L_x\frac{\partial
u_{-x}^{(2)}}{\partial\eta}+L_x\frac{\partial}{\partial\eta}(n_-^{(1)}u_{-x}^{(1)})+L_y\frac{\partial
u_{-y}^{(2)}}{\partial\eta}\nonumber\\
&&\hspace*{-6mm}+L_y\frac{\partial}{\partial\eta}(n_-^{(1)}u_{-y}^{(1)})+L_z\frac{\partial
u_{-z}^{(3)}}{\partial\eta}
+L_z\frac{\partial}{\partial\eta}(n_-^{(1)}u_{-z}^{(2)})+L_z\frac{\partial}{\partial\eta}(n_-^{(2)}u_{-z}^{(1)})=0,
\label{C7b}\\
&&\hspace*{12mm}\frac{\partial u_{+z}^{(1)}}{\partial
T}-V_p\frac{\partial
u_{+z}^{(3)}}{\partial\eta}+L_z\frac{\partial}{\partial\eta}(u_{+z}^{(1)}u_{+z}^{(2)})+L_z\frac{\partial\phi^{(3)}}{\partial\eta}=0,
\label{C7c}\\
&&\hspace*{12mm}\frac{\partial u_{-z}^{(1)}}{\partial
T}-V_p\frac{\partial
u_{-z}^{(3)}}{\partial\eta}+L_z\frac{\partial}{\partial\eta}(u_{-z}^{(1)}u_{-z}^{(2)})-L_z\beta\frac{\partial\phi^{(3)}}{\partial\eta}=0,
\label{C7d}\\
&&\hspace*{16mm}L_z\frac{\partial\phi^{(3)}}{\partial\eta}-K_{11}L_z\frac{\partial
n_e^{(3)}}{\partial\eta}-F\frac{\partial}{\partial\eta}(n_e^{(1)}n_e^{(2)})=0,
\label{C7e}\\
&&\hspace*{22mm}\frac{\partial^2\phi^{(1)}}{\partial\eta^2}=\delta
n_e^{(3)}+\sigma n_-^{(3)}-n_+^{(3)}, \label{C7f}
\end{eqnarray}

where $F=K_{11}L_z(\gamma-2)$

Now combining these higher order equations together with
Eqs.(\ref{C7a})-(\ref{C7f}) one can obtain

\begin{eqnarray}
&&\hspace*{-10mm}\frac{\partial\phi^{(1)}}{\partial T} +
M\phi^{(1)2} \frac{\partial \phi^{(1)}}{\partial
\eta}+N\frac{\partial^3 \phi^{(1)}}{\partial \eta^3}=0,
\label{DIAK-dV}
\end{eqnarray}

This is well-known mK-dV equation that describes the obliquely
propagating IA waves in a magnetized multi-ion quantum plasma.
where $M$ and $N$ are given by

\begin{eqnarray}
&&\hspace*{-10mm}M=\frac{K_{11}V_p}{2\delta}\left[\frac{5L_z^6\beta^3\sigma}{6V_p^6}+\frac{5L_z^6}{6V_p^6}-\frac{\delta(\gamma-2)^2}{9K_{11}^3}\right],
\label{B5a}\\
&&\hspace*{-10mm}N=\frac{K_{11}V_p}{2\delta}\left[1+\frac{\sigma\beta(1-L_z^2)}{\omega_{c-}^2}+\frac{(1-L_z^2)}{\omega_{c+}^2}\right]
\label{B5b}
\end{eqnarray}

The stationary solitary wave solution of the standard mK-dV
equation is obtained by considering a frame $\xi=\eta-u_{0}T$
(moving with speed $u_{0}$) and the solution is,

\begin{eqnarray}
{\rm \phi^{(1)}}=\rm \phi_m{\rm[sech}(\frac{\xi}{\varpi})],
\label{solK-dV}
\end{eqnarray}

\noindent where the amplitude, $\phi_m=\sqrt{(6u_{0}/M)}$ and the
width $\varpi= {\sqrt{(N/u_{0})}}$.

\begin{figure}[t!]
\centering
\includegraphics[width=5.2cm]{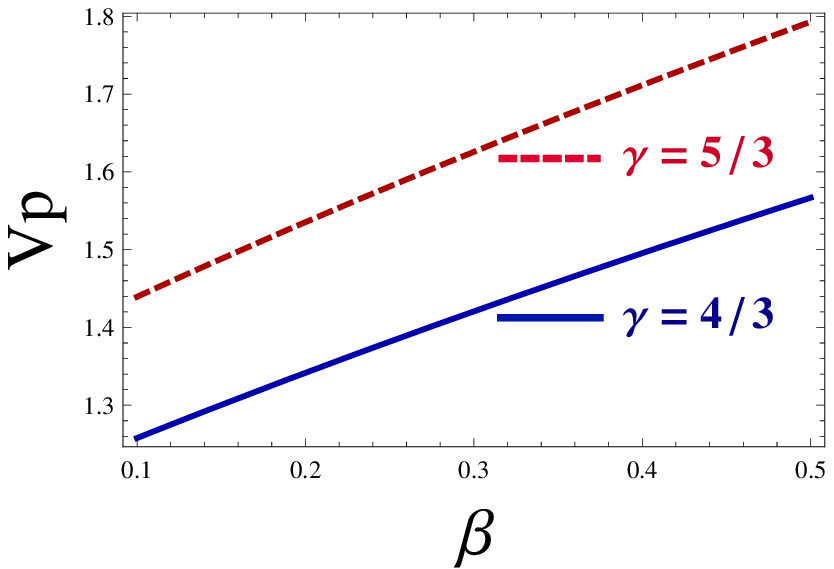}\hspace{1cm}
\includegraphics[width=5.2cm]{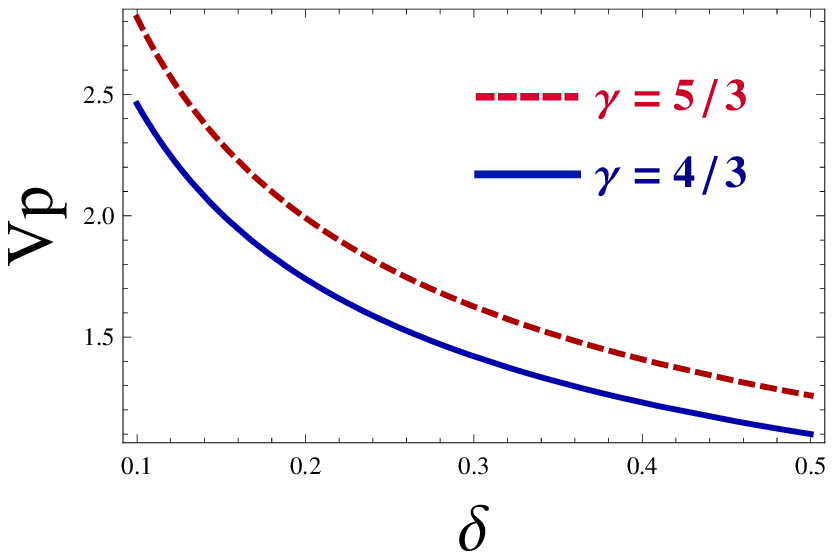}

\Large{(a)\quad\,\,\,\,\,\quad\,\,\quad\,\,\,\,\,\quad\,\quad\,\,\,\,\,\quad\,\,\,\,\,\,\,\,(b)}\vspace{1cm}

\includegraphics[width=5.5cm]{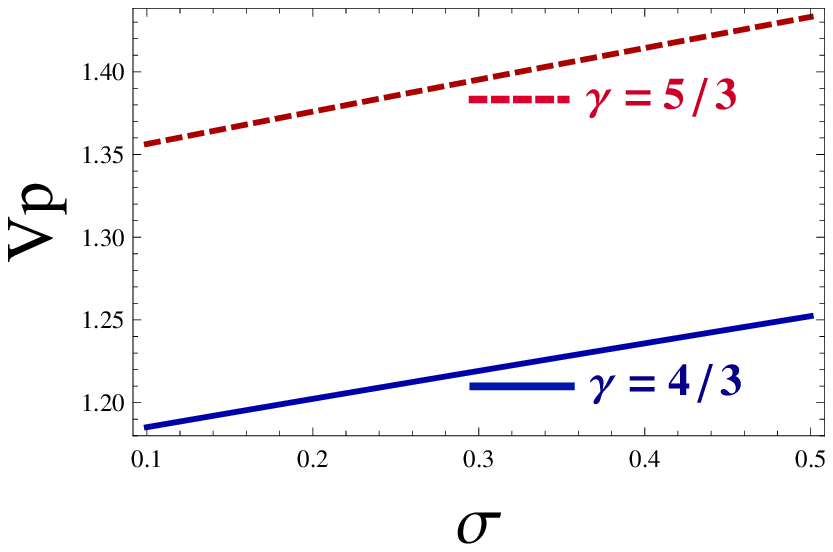}

\Large{(c)} \caption{(Color online) Showing the variation of
phase speed $V_p$ with (a) the ratio of the masses of the negative
and the positive ion multiplied by their charge per ion $\beta$,
(b) the ratio of the number density of electrons and positive ions
multiplied by charge per positive ion $\delta$, and (c) the ratio
of number density of negative and positive ions multiplied by
their charge per ion $\sigma$, for $\theta=10^0$. The red dashed
line represents the non-relativistic case and the blue solid line
represents the ultra-relativistic case.}
 \label{Fig1}
\end{figure}

\begin{figure}[!t]
\centering
\includegraphics[width=6.0cm]{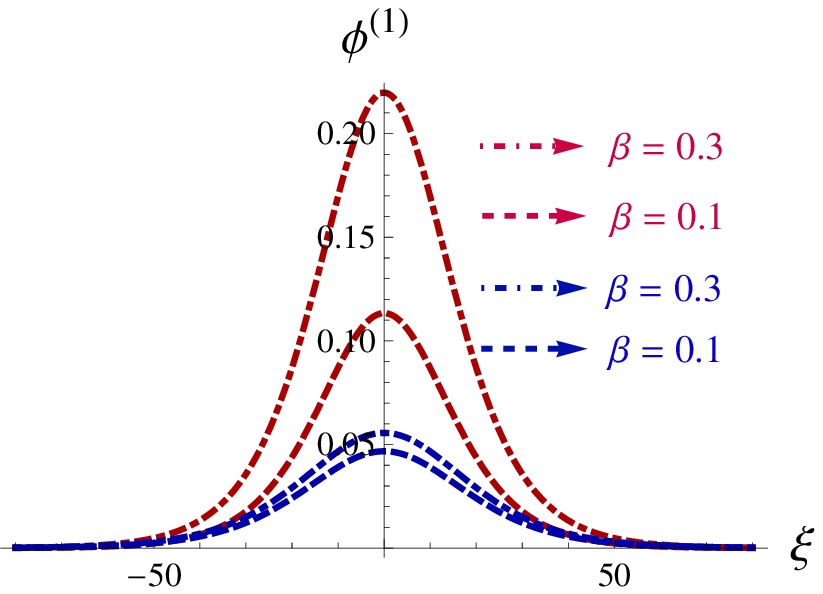}\hspace{1cm}
\includegraphics[width=5.2cm]{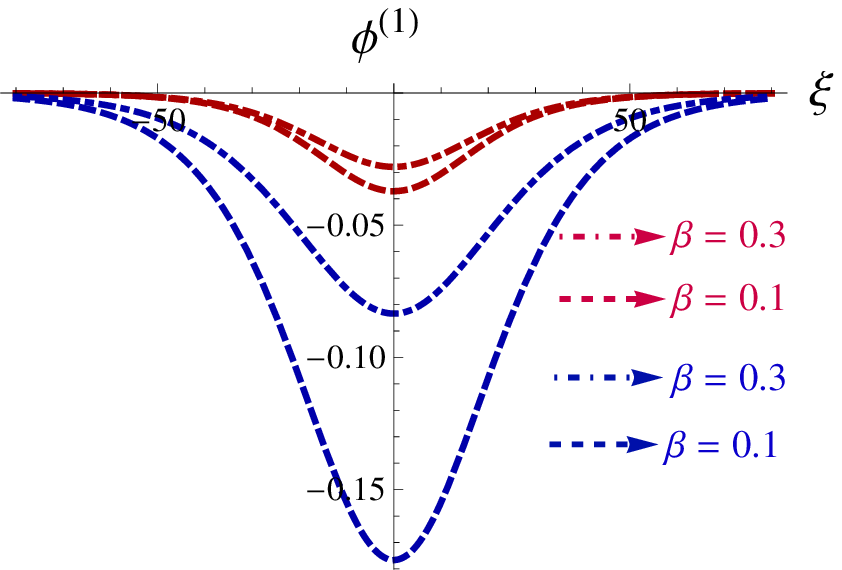}

\Large{(a)\,\,\,\,\,\,\,\,\,\,\,\,\,\,\,\,\,\,\,\,\,\,\,\,\,\,\,\,\,\,\,\,\,\,\,\,\,\,\,\,\,\,\,\,\,\,\,\,\,\,\,\,\,\,\,\,\,\,\,\,\,\,\,\,\,\,\,\,\,\,\,\,\,\,\,\,\,(b)}\vspace{1cm}

 \caption{(Color online) Showing the variation of the positive and
 negative potential K-dV solitons $\phi^{(1)}$ with $\beta$ for $u_0=0.01$,
$\omega_{ci}=0.5$, $\delta=0.3$, and $\theta=8^0$ in case of both
non-relativistic and ultra-relativistic limit, where (a) for
$\sigma<\sigma_c$ and (b) for $\sigma>\sigma_c$. The blue line
represents the non-relativistic case and the red line represents
the ultra-relativistic case.}
 \label{Fig2}
\end{figure}

\begin{figure}[!t]
\centering
\includegraphics[width=5.2cm]{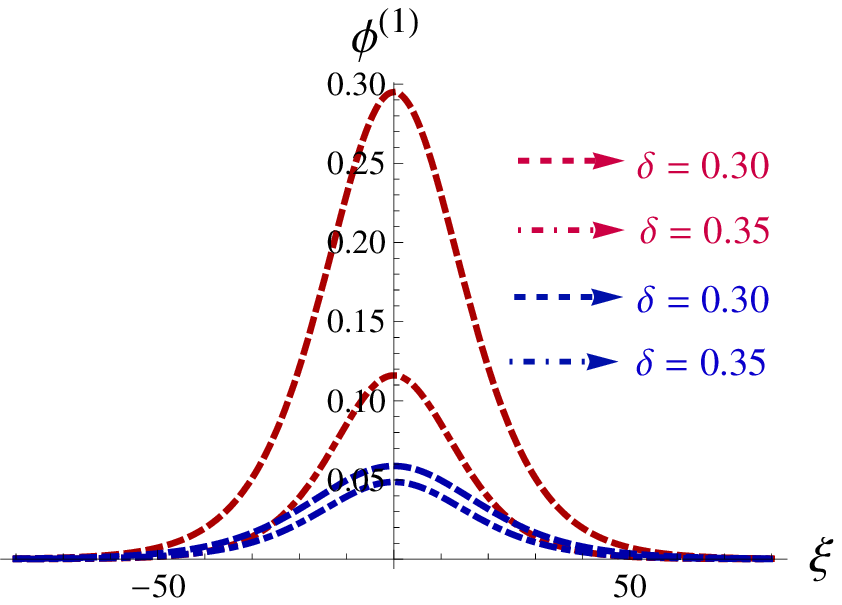}\hspace{1cm}
\includegraphics[width=5.2cm]{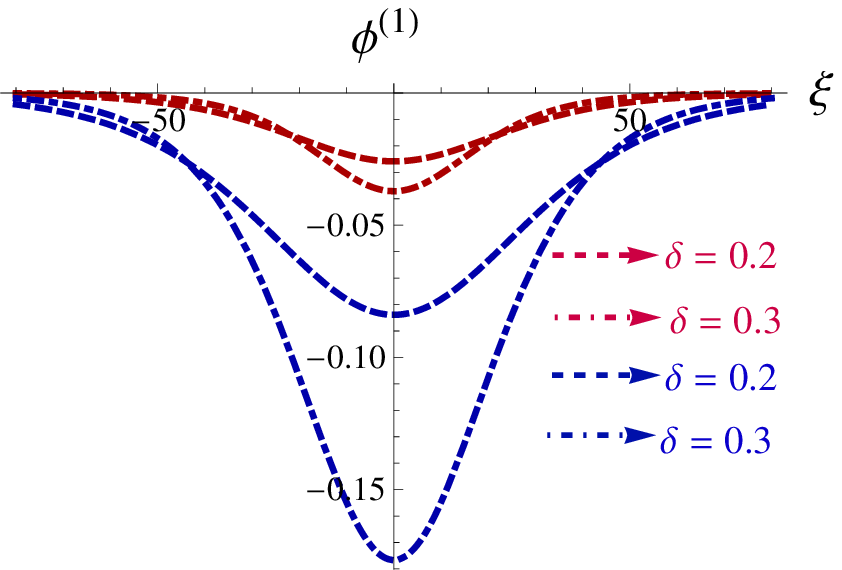}

\Large{(a)\,\,\,\,\,\,\,\,\,\,\,\,\,\,\,\,\,\,\,\,\,\,\,\,\,\,\,\,\,\,\,\,\,\,\,\,\,\,\,\,\,\,\,\,\,\,\,\,\,\,\,\,\,\,\,\,\,\,\,\,\,\,\,\,\,\,\,\,\,\,\,\,\,\,\,\,\,(b)}\vspace{1cm}

 \caption{(Color online) Showing the variation of the positive and
 negative potential K-dV solitons $\phi^{(1)}$ with $\delta$ for $u_0=0.01$,
$\omega_{ci}=0.5$, $\beta=0.3$, and $\theta=8^0$ in case of both
non-relativistic and ultra-relativistic limit, where (a) for
$\sigma<\sigma_c$ and (b) for $\sigma>\sigma_c$. The Blue line
represents the non-relativistic case and the Red line represents
the ultra-relativistic case.}
 \label{Fig3}
\end{figure}

\begin{figure}[!t]
\centering
\includegraphics[width=6.2cm]{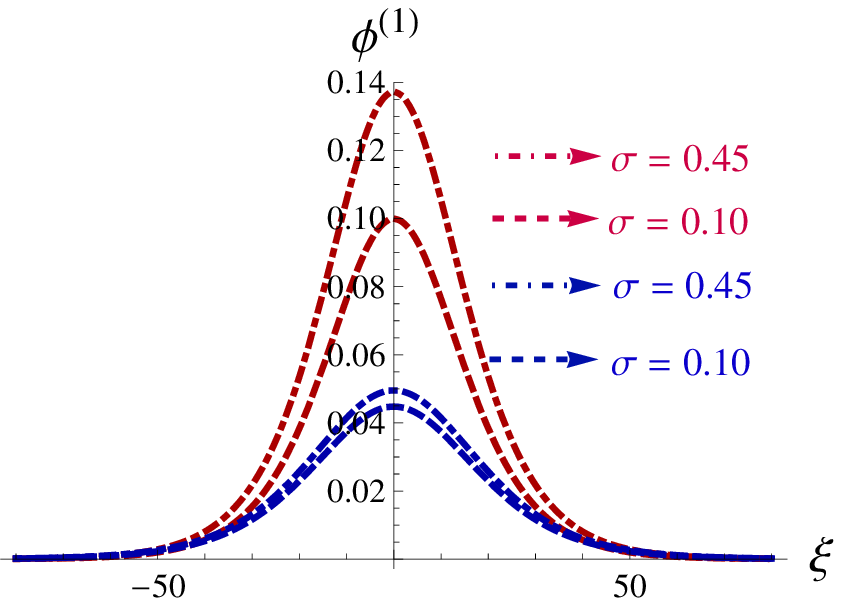}\hspace{1cm}
\includegraphics[width=6.2cm]{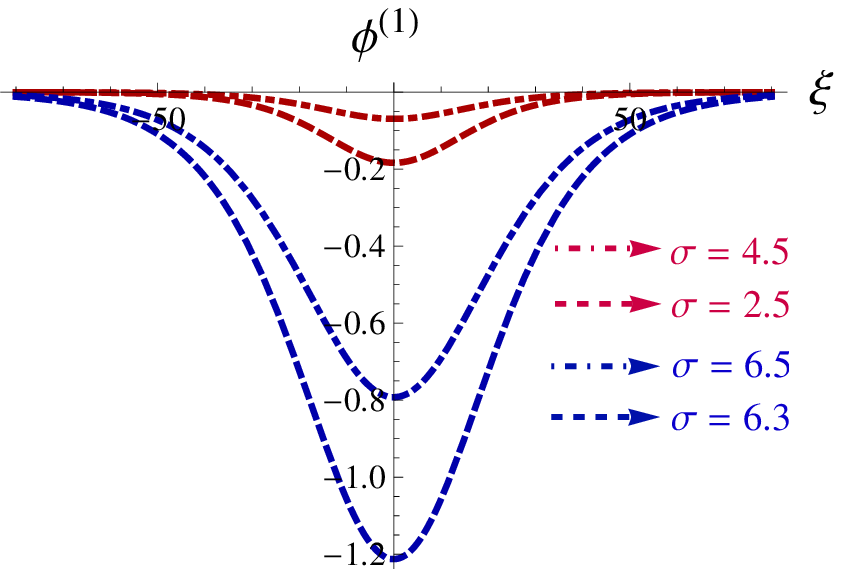}

\Large{(a)\,\,\,\,\,\,\,\,\,\,\,\,\,\,\,\,\,\,\,\,\,\,\,\,\,\,\,\,\,\,\,\,\,\,\,\,\,\,\,\,\,\,\,\,\,\,\,\,\,\,\,\,\,\,\,\,\,\,\,\,\,\,\,\,\,\,\,\,\,\,\,\,\,\,\,\,\,(b)}\vspace{1cm}

 \caption{(Color online)(a) Showing the amplitude variation of the K-dV solitons $\phi^{(1)}$ with $\sigma$ for $u_0=0.01$,
$\omega_{ci}=0.5$, and $\theta=8^0$ in case non-relativistic
limit and (b) Showing the contour plot of K-dV solitons potential
$\phi^{(1)}$ with $\sigma$ for $u_0=0.01$, $\omega_{ci}=0.5$, and
$\theta=8^0$ in case non-relativistic limit.}
 \label{Fig4}
\end{figure}

\begin{figure}[!t]
\centering
\includegraphics[width=5.2cm]{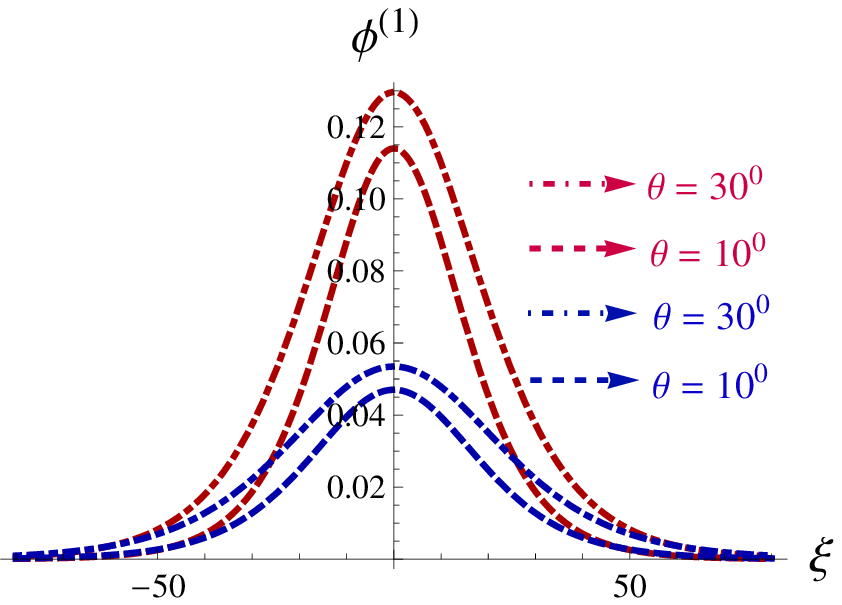}\hspace{1cm}
\includegraphics[width=5.2cm]{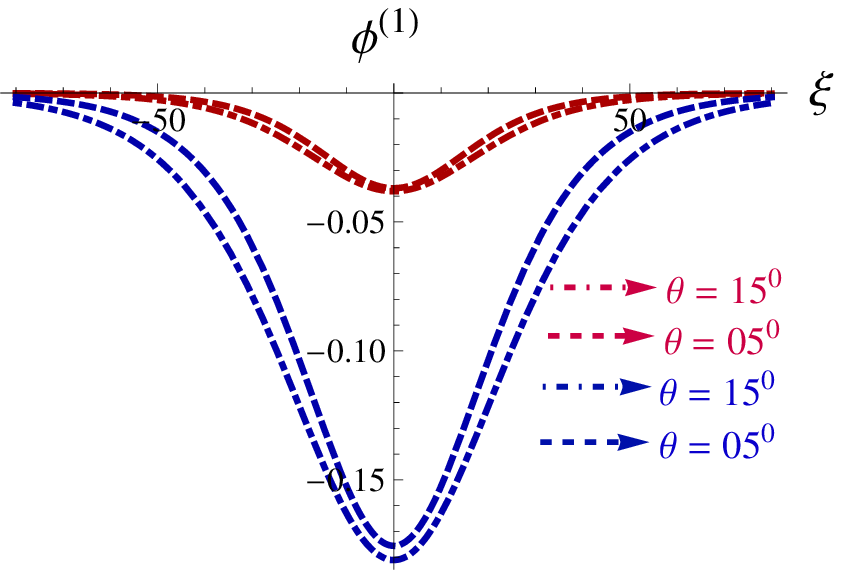}

\Large{(a)\,\,\,\,\,\,\,\,\,\,\,\,\,\,\,\,\,\,\,\,\,\,\,\,\,\,\,\,\,\,\,\,\,\,\,\,\,\,\,\,\,\,\,\,\,\,\,\,\,\,\,\,\,\,\,\,\,\,\,\,\,\,\,\,\,\,\,\,\,\,\,\,\,\,\,\,\,(b)}\vspace{1cm}

 \caption{(Color online) Showing the variation of the positive and
 negative potential K-dV solitons $\phi^{(1)}$ with $\theta$ for $u_0=0.01$,
$\omega_{ci}=0.5$, $\beta=0.3$ and $\delta=0.3$ in case of both
non-relativistic and ultra-relativistic limit, where (a) for
$\sigma<\sigma_c$ and (b) for $\sigma>\sigma_c$. The Blue line
represents the non-relativistic case and the Red line represents
the ultra-relativistic case.}
 \label{Fig5}
\end{figure}

\section{DISCUSSION AND RESULTS}

The propagation of IASWs in a magnetized plasma containing of
both non-relativistic and ultra-relativistic degenerate
electrons, non-relativistic degenerate inertial ions of both
positively and negatively charged, and positively charged
immobile heavy elements has been studied numerically. To drive
K-dV and mK-dV equation we used the well-known reductive
perturbation method and then we have studied and analyzed the
IASWs solution. We observed and analyzed that both compressive
and rarefactive solitary waves (SWs) are found to exist. We have
investigate the effects of the different intrinsic parameters
(namely the ratio of the masses of the negative and the positive
ion multiplied by their charge per ion $\beta$, the ratio of the
number density of electrons and positive ions multiplied by charge
per positive ions $\delta$, the ratio of number density of
negative and positive ion multiplied by their charge per ion
$\sigma$, obliqueness $\theta$, cyclotron frequency
$\omega_{ci}$, relativistic factor ) on the dynamic properties of
IASWs. The amplitude of SWs has been modified by the degenerate
pressure of electrons illustrated from the non-relativistic
$({P_e}\propto{n_e}^{5/3})$ to ultra-relativistic
$({P_e}\propto{n_e}^{4/3})$ regime. Adopting Chandrasekhar's
equation of state for relativistically degenerate electrons, it
has been examined that the relativistic factor greatly affects
the speed of IASWs where for the nonrelativistic degenerate
electrons, $\gamma=\frac{5}{3}$ and for the ultrarelativistic
degenerate electrons, $\gamma=\frac{4}{3}$, and thus found that
the relativistic factor, $\gamma=5/3 ~(nonrelativistic)
> \gamma=4/3 ~(ultrarelativistic)$ in every cases.

Here we have numerically obtained that for $\lambda=0$, the
amplitude of the K-dV solitons become infinitely large, and the
K-dV solution is no longer valid at $\lambda\simeq0$. It has been
observed that the solution of the K-dV equation supports both
compressive (positive) and rarefactive (negative) structures
depending on the critical value of $\sigma$. In our present
investigation, we have found that for $\sigma_c=1.70592$, the
amplitude of the SWs breaks down due to the vanishing of the
nonlinear coefficient $\lambda$. We have observed that at
$\sigma<1.70592$, positive (compressive) potential SWs exist,
whereas at $\sigma>1.70592$, negative (rarefactive) SWs exist
(shown in Figs. 2-4).

\begin{figure}[t!]
\centering
\includegraphics[width=5.2cm]{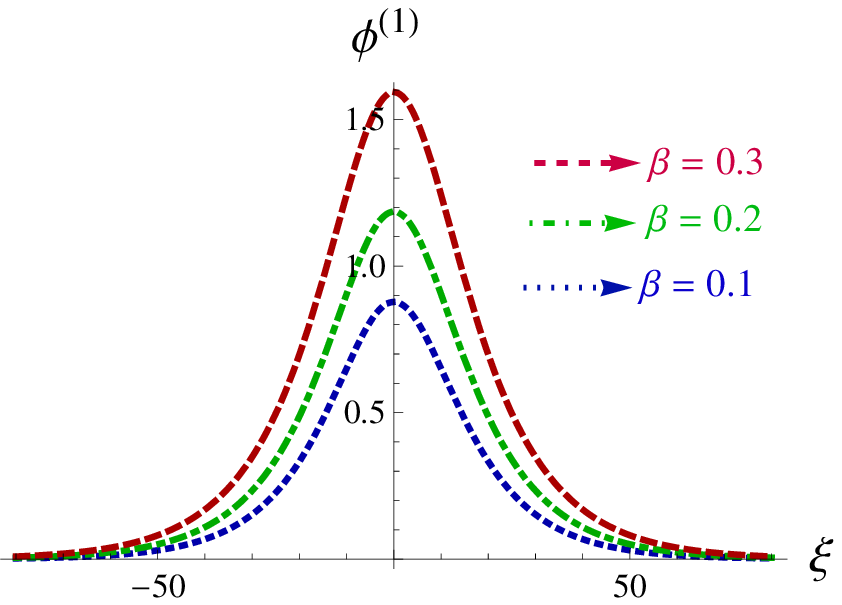}\hspace{1cm}
\includegraphics[width=5.2cm]{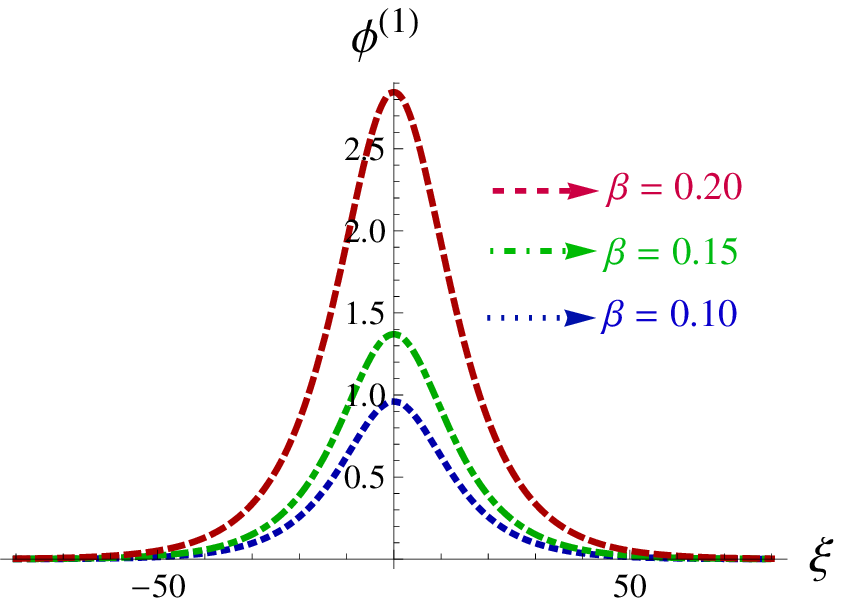}

\Large{(a)\,\,\,\,\,\,\,\,\,\,\,\,\,\,\,\,\,\,\,\,\,\,\,\,\,\,\,\,\,\,\,\,\,\,\,\,\,\,\,\,\,\,\,\,\,\,\,\,\,\,\,\,\,\,\,\,\,\,\,\,\,\,\,\,\,\,\,\,\,\,\,\,\,\,\,\,\,(b)}\vspace{1cm}

\caption{(Color online) Showing the variation of the amplitude of
magnetized mK-dV solitons with $\beta$, where (a) for
non-relativistic limit and (b) for ultra-relativistic limit. The
other plasma parameters are kept fixed.} \label{6}
\end{figure}

\begin{figure}[t!]
\centering
\includegraphics[width=6.2cm]{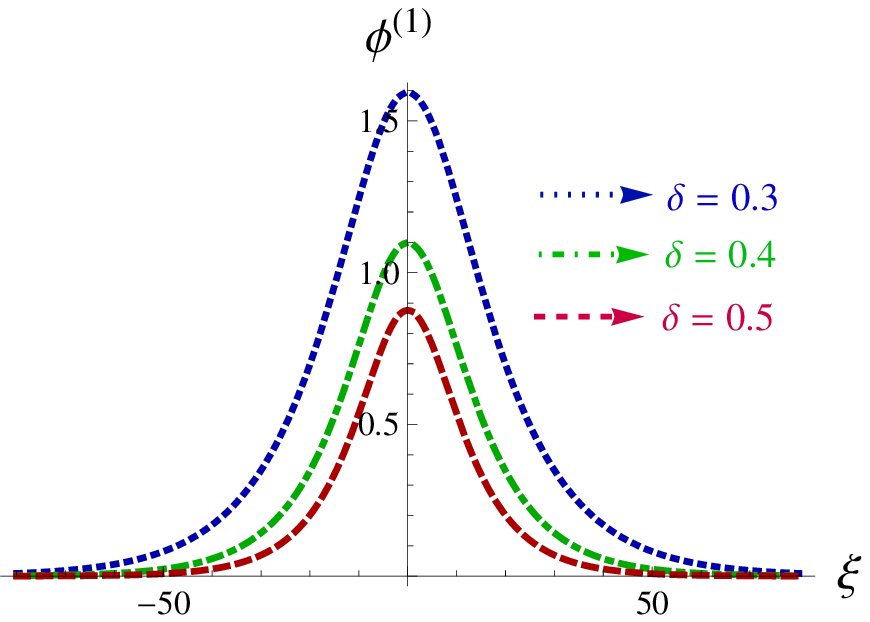}\hspace{1cm}
\includegraphics[width=6.2cm]{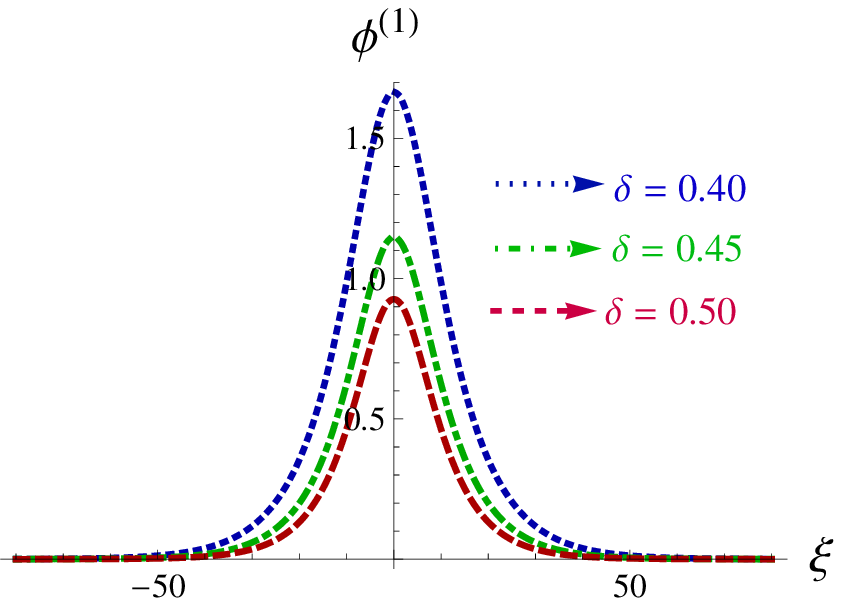}

\Large{(a)\,\,\,\,\,\,\,\,\,\,\,\,\,\,\,\,\,\,\,\,\,\,\,\,\,\,\,\,\,\,\,\,\,\,\,\,\,\,\,\,\,\,\,\,\,\,\,\,\,\,\,\,\,\,\,\,\,\,\,\,\,\,\,\,\,\,\,\,\,\,\,\,\,\,\,\,\,(b)}\vspace{1cm}

\caption{(Color online) (a) Showing the amplitude variation of
the mK-dV solitons $\phi^{(1)}$ with $\delta$ for non-relativistic
limit and (b) Showing the contour plot of mK-dV solitons potential
$\phi^{(1)}$ with $\delta$ for non-relativistic limit.} \label{7}
\end{figure}

\begin{figure}[t!]
\centering
\includegraphics[width=5.2cm]{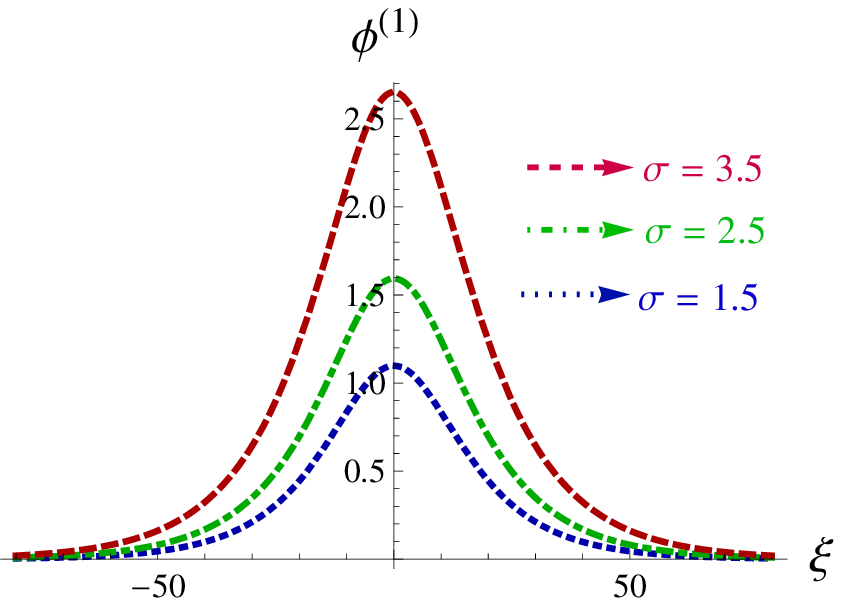}\hspace{1cm}
\includegraphics[width=5.2cm]{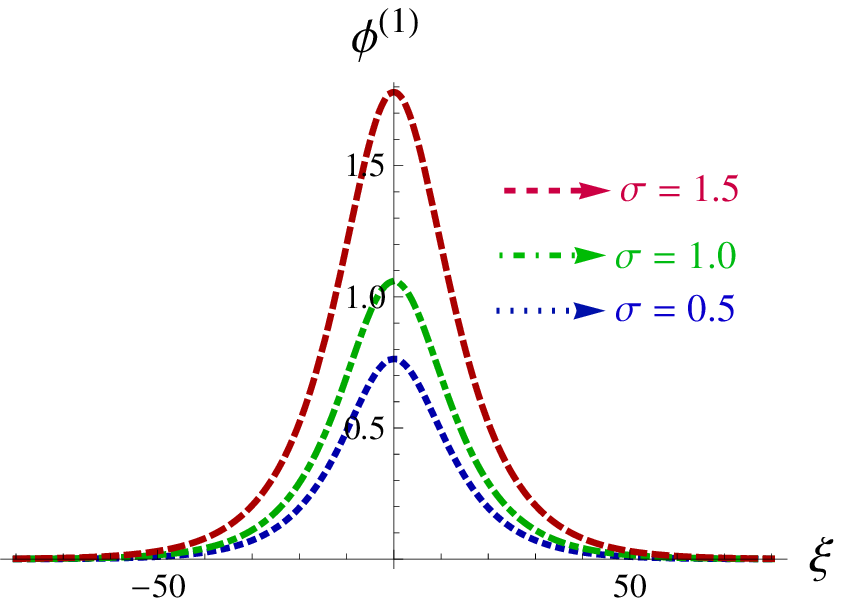}

\Large{(a)\,\,\,\,\,\,\,\,\,\,\,\,\,\,\,\,\,\,\,\,\,\,\,\,\,\,\,\,\,\,\,\,\,\,\,\,\,\,\,\,\,\,\,\,\,\,\,\,\,\,\,\,\,\,\,\,\,\,\,\,\,\,\,\,\,\,\,\,\,\,\,\,\,\,\,\,\,(b)}\vspace{1cm}

\caption{(Color online) Showing the variation of the amplitude of
magnetized mK-dV solitons with $\sigma$, where (a) for
non-relativistic limit and (b) for ultra-relativistic limit. The
other plasma parameters are kept fixed.} \label{8}
\end{figure}

\begin{figure}[t!]
\centering
\includegraphics[width=5.2cm]{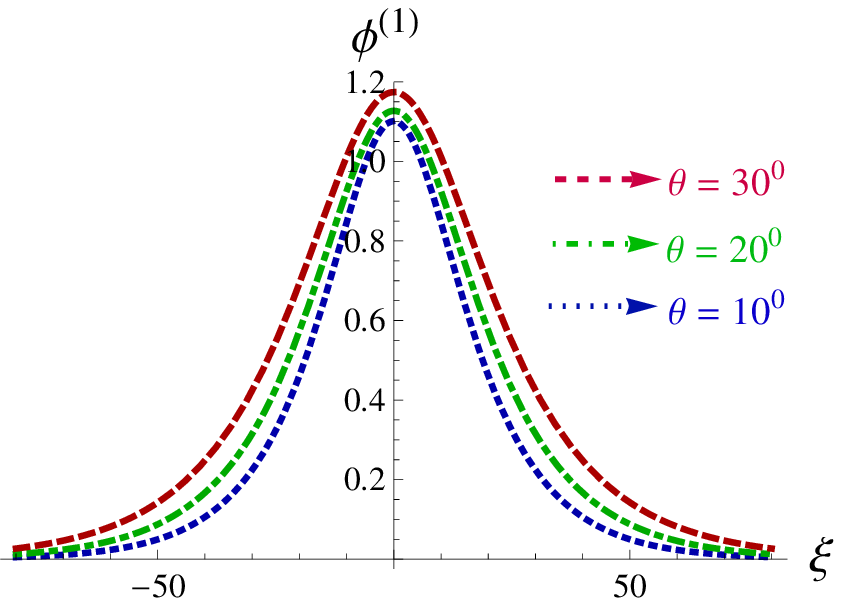}\hspace{1cm}
\includegraphics[width=5.2cm]{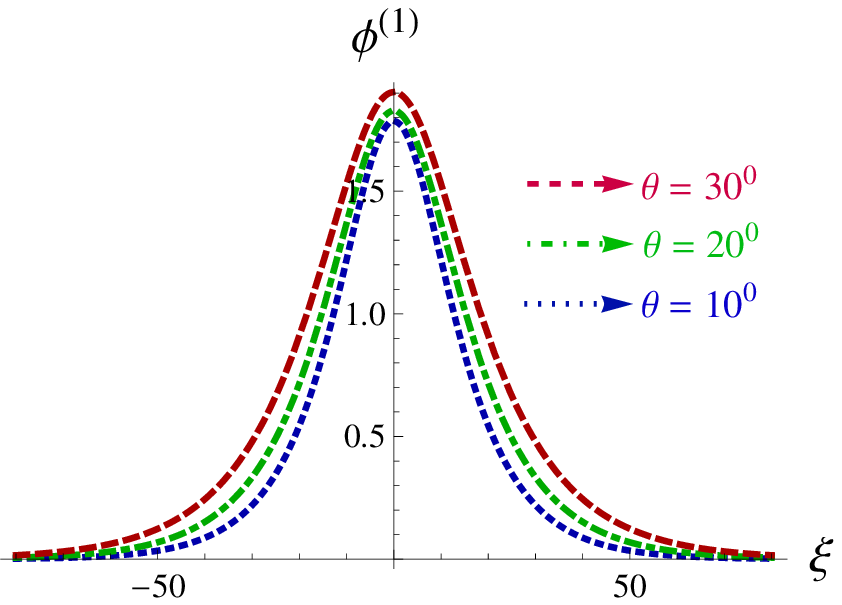}

\Large{(a)\,\,\,\,\,\,\,\,\,\,\,\,\,\,\,\,\,\,\,\,\,\,\,\,\,\,\,\,\,\,\,\,\,\,\,\,\,\,\,\,\,\,\,\,\,\,\,\,\,\,\,\,\,\,\,\,\,\,\,\,\,\,\,\,\,\,\,\,\,\,\,\,\,\,\,\,\,(b)}\vspace{1cm}

\caption{(Color online) Showing the variation of the amplitude of
magnetized mK-dV solitons with $\theta$, (a) for non-relativistic
limit and (b) for ultra-relativistic limit. The other plasma
parameters are kept fixed.} \label{9}
\end{figure}

Figure 1 shows the variation of phase speed $V_p$ with (a) the
ratio of the masses of the negative and the positive ion
multiplied by their charge per ion $\beta$. We observed that the
phase speed $V_p$ increases with the increasing values of
$\beta$. (b) the ratio of the number density of electron and
positive ion multiplied by charge per positive ion $\delta$. It
is shown that with the increase of $\delta$ the phase speed $V_p$
decreasing gradually. (c) the ratio of number density of negative
and positive ions multiplied by their charge per ion $\sigma$. We
found that the phase speed $V_p$ increases with the increasing
values of $\sigma$. These outcomes are also clear from the phase
speed equation of our considered model. It is obvious that the
phase speed is always higher for non-relativistic case than those
for the ultra-relativistic case. It is due to the variation of the
values of the relativistic $\gamma$ factor which is described in
introduction.

The effect of $\beta$ on the amplitude of K-dV soliton has shown
in Figure 2. It is shown that the amplitude and width of K-dV
soliton increases with the increasing values of $\beta$ for both
non-relativistic and ultra-relativistic case. Actually, this
happens because it increases both the dispersive coefficient and
the nonlinearity coefficient. We also found that for positive K-dV
soliton $\sigma<\sigma_c$ and for negative K-dV soliton
$\sigma>\sigma_c$.

Figure 3 shows the effect of the variation of $\delta$ with the
amplitude of K-dV soliton. It is observed that the solitary
profile decreases with increasing values of $\delta$. The effect
of $\sigma$ on the amplitude of K-dV soliton was shown in Figure
4. Here we observed that the K-dV soliton increases with the
increasing values of $\sigma$. Physically, as the total value of
dispersive coefficent increases (i.e., increasing in dispersion of
the system), the potential of the solitons increases.

It is shown that the width of the K-dV solitary profiles decrease
with the increasing values of $\omega_{ci}$ both for
non-relativistic and ultra-relativistic cases. We also observed
that the width of the K-dV solitary profile is higher when the
plasma system being non-relativistic degenerate case than the
ultra-relativistic degenerate case shown in Figure 5 where the
width goes to zero when the obliqueness tends to zero and not
valid for $\theta>90^0$.

Figure 6 shows the effect of obliqueness of the propagation
direction, as expressed via $\theta$, is observed for
relativistically degenerate electrons, considering $\gamma=5/3$
and $\gamma=4/3$. The variation of amplitude of IASWs is taken
place with different values of obliqueness of the wave
propagation. It is seen that the amplitude of the magnetized K-dV
and mK-dV solitons increases with the increasing of obliqueness,
i.e., the angle ($\theta$) between the direction of wave
propagation and the magnetic field, $B_0$. It is seen that as the
value of $\theta$ increases, the amplitude of the solitary waves
increases, while their width increases for the lower range of
$\delta$ (from $0^{\circ}$ to about $55^{\circ}$), and decreases
for its higher range (from $55^{\circ}$ to about $90^{\circ}$).
As $\delta \rightarrow90^{\circ}$, the width goes to $0$, and the
amplitude goes to $\infty$. It is likely that for large angles,
the assumption that the waves are electrostatic is no longer
valid, and we should look for fully electromagnetic structures.
Our present investigation is only valid for small value of
$\delta$ but invalid for arbitrary large value of $\theta$. In
case of larger values of $\theta$, the wave amplitude becomes
large enough to break the validity of the reductive perturbation
method.

For mK-dV solitons, only compressive solitons are exists which is
well-known in plasma literature. It is observed that like K-dV
soliton the amplitude of magnetized mK-dV solitons increase with
with the increasing of $\beta$ shown in Figure 6. This occurs
because $\beta$ increases both the dispersive coefficient and the
nonlinearity coefficient. It is also investigated that the
amplitude of the mK-dV solitary profiles decreases with the
increasing values of $\delta$ as K-dV solitons shown in Figure 7.
From Figure 8 we found that for both non-relativistic and
ultra-relativistic cases the amplitude of mK-dV solitons increase
with the increasing values of $\sigma$. From Figure 9 we found
that with the increase in obliqueness the amplitude of magnetized
mK-dV solitons increase like K-dV soliton. Actually, the
increasing or decreasing of physical parameters strongly
responsible for the variation of both the dispersive coefficient
(produce due to the dispersion) and the nonlinearity coefficient
which makes the solitary strength high or low.

In conclusion, the results of the present investigation should be
useful for understanding the nonlinear features of IASWs in a
relativistic degenerate quantum multi-ion plasmas which are found
in a number of astrophysical plasma systems, such as, neutron
stars, white dwarfs, etc. where presence of heavy elements,
obliqueness of wave propagation, and relativistic degenerate
electrons play a crucial role.
\\\


\begin{thebibliography}{100}

\bibitem{Tiwari1980} R.S. Tiwari, S.R. Sharma, Modulation instability of ion-acoustic waves in a multi-ion plasma, Phys. Lett. A 77 (1980) 30.

\bibitem{D'Angelo1990} N. D'Angelo, Low-frequency electrostatic waves in dusty plasmas, Planet. Space Sci. 38 (1990) 1143.

\bibitem{Chen2001} Chen Yin-Hua, Lu Wei, Wang Wen-Hao, The nonlinear langmuir waves in a multi-ion-component plasma, Commun. Theor. Phys. 35 (2001) 223.

\bibitem{Nakamura2003} Y. Nakamura, Y. Saitou, Observation of ion-acoustic waves in two-ion-species plasmas, Plasma Phys. Controlled Fusion 45 (2003) 759.

\bibitem{Mamun2009} A.A. Mamun, R.A. Cairns, P.K. Shukla, Dust negative ion acoustic shock waves in a dusty multi-ion plasma, Phys. Lett. A 373 (2009) 2355.

\bibitem{Perrone2013}  D. Perrone, F. Valentini, S. Servidio, S. Dalena, P. Veltri, Vlasov simulations of multi-ion plasma turbulence in the solar wind, Astrophys. J. 762 (2013) 99.

\bibitem{Shahmansouri2015} M. Shahmansouri, H. Alinejad, M. Tribeche, Multi-ion double layers in a magnetized plasma, Commun. Theor. Phys. 64 (2015) 555.

\bibitem{Rizzato1987} F.B. Rizzato, R.S. Schneider, D. Dillenburg, Temperature effects on ion acoustic solitons in plasmas with near critical density of negative ions, Plasma Phys. Control. Fusion 29 (1987) 1127.

\bibitem{Gottscho1986} R.A. Gottscho, C.E. Gaebe, Negative ion kinetics in RF glow discharges, IEEE Trans. Plasma Sci. 14 (1986) 92.

\bibitem{Bacal1979} M. Bacal, G.W. Hamilton, H- and D- Production in Plasmas, Phys. Rev. Lett. 42 (1979) 1538.

\bibitem{Jacquinot1977} J. Jacquinot, B.D. McVey, J.E. Scharer, Mode conversion of the fast magnetosonic wave in a deuterium-hydrogen tokamak plasma, Phys. Rev. Lett. 39 (1977) 88.

\bibitem{Ortner2016} A. Ortner, D. Schumacher, W. Cayzac, A. Frank, M.M. Basko, S. Bedacht, A. Blazevic, S. Faik, D. Kraus, T. Rienecker, G.
Schaumann, An. Tauschwitz, F. Wagner and M. Roth, A novel
experimental setup for energy loss and charge state measurements
in dense moderately coupled plasma using laser-heated hohlraum
targets, J. Phys.: Conf. Ser. 688 (2016) 012081.


\bibitem{Washimi1966} H. Washimi, T. Taniuti, Propagation of ion-acoustic solitary waves of small amplitude, Phys. Rev. Lett. 17 (1966) 996.

\bibitem{Sagdeev1966} R.Z. Sagdeev, Cooperative phenomena and shock waves in collisionless plasmas, Rev. Plasma Phys. 4 (1966) 23.

\bibitem{Ikezi1970} H. Ikezi, R.J. Taylor, D.R. Baker, Formation and interaction of ion-acoustic solitions, Phys. Rev. Lett. 25 (1970) 11.

\bibitem{Mase1975} A. Mase, T. Tsukishima, Measurements of ion wave turbulence by microwave scattering, Phys. Fluids 18 (1975) 464.

\bibitem{Baboolal1990} S. Baboolal, R. Bharuthram, Cut-off conditions and existence domains for large-amplitude ion-acoustic solitons and double layers in fluid plasmas, J. Plasma Phys. 44 (1990) 1.

\bibitem{Bharuthram1992} R. Bharuthram, P.K. Shukla, Large amplitude ion-acoustic solitons in a dusty plasma, Planet. Space Sci. 40 (1992) 973.

\bibitem{Popel1995} S.I. Popel, S.V. Vladimirov, P.K. Shukla, Ion-acoustic solitons in electron–positron–ion plasmas, Phys. Plasmas 2 (1995) 716.

\bibitem{Mamun1997} A.A. Mamun, Effects of ion temperature on electrostatic solitary structures in nonthermal plasmas, Phys. Rev. E 55 (1997) 1852.

\bibitem{Mamun2001} A.A. Mamun, P.K. Shukla, Spherical and cylindrical dust acoustic solitary waves. Phys. Lett. A 290 (2001) 173.

\bibitem{Shahmansouri2014} M. Shahmansouri, M. Tribeche, Propagation properties of ion acoustic waves in a magnetized superthermal bi-ion plasma, Astrophys. Space Sci. 350 (2014) 781.

\bibitem{Baboolal1989} S. Baboolal, R. Bharuthram, M.A. Hellberg, Arbitrary-amplitude theory of ion-acoustic solitons in warm multi-fluid plasmas, J. Plasma Phys. 41 (1989) 341.

\bibitem{Rice1993} W.K.M. Rice, M.A. Hellberg, R.L. Mace,  S. Baboolal, Finite electron mass effects on ion-acoustic solitons in a two electron temperature plasma, Phys. Lett. A 174 (1993) 416.

\bibitem{Ghosh1996} S.S. Ghosh, K.K. Ghosh, A.N. Sekar Iyengar, Large Mach number ion acoustic rarefactive solitary waves for a two electron temperature warm ion plasma, Phys. Plasmas 3 (1996) 3939.

\bibitem{Shah2015d} M.G. Shah, M.M. Rahman, M.R. Hossen, A.A. Mamun, Roles of superthermal electrons and adiabatic heavy ions on heavy-ion-acoustic solitary and shock waves in a multi-component plasma, Commun. Theor. Phys. 64 (2015) 208.

\bibitem{Shah2016} M.G. Shah, M.M. Rahman, M.R. Hossen, A.A. Mamun, Properties of cylindrical and spherical heavy ion-acoustic solitary and shock structures in a multi-species plasma with superthermal electrons, Plasma Phys. Rep. 42 (2016) 168.

\bibitem{Alinejad2011} H. Alinejad, A.A. Mamun, Oblique propagation of electrostatic waves in a magnetized electron-positron-ion plasma with superthermal electrons, Phys. Plasmas 18 (2011) 112103.

\bibitem{Hossen2014} M.R. Hossen, L. Nahar, S. Sultana, A.A. Mamun, Roles of positively charged heavy ions and degenerate plasma pressure on cylindrical and spherical ion acoustic solitary waves, Astrophys. Space Sci. 353 (2014) 123.

\bibitem{Shah2015c} M.G. Shah, M.R. Hossen, S. Sultana, A.A. Mamun, Positron-acoustic shock waves in a degenerate multi-component plasma, Chin. Phys. Lett. 32 (2015) 085203.

\bibitem{Yu1980} M.Y. Yu, P.K. Shukla, S. Bujarbarua, Fully nonlinear ion-acoustic solitary waves in a magnetized plasma, Phys. Fluids 23 (1980) 2146.

\bibitem{Shukla2001} P.K. Shukla, A.A. Mamun, Dust-acoustic shocks in a strongly coupled dusty plasma, IEEE Trans. Plasma Sci. 29 (2001) 221.

\bibitem{Mamun1998} A.A. Mamun, M.N. Alam, A.K. Das, Z. Ahmed, T.K. Datta, Obliquely propagating electrostatic solitary structures in a hot magnetized dusty plasma, Phys. Scr. 58 (1998) 72.

\bibitem{Sultana2012} S. Sultana, I. Kourakis, M.A. Hellberg, Oblique propagation of arbitrary amplitude electron acoustic solitary waves in magnetized kappa-distributed plasmas, Plasma Phys. Controlled Fusion 54 (2012) 105016.


\bibitem{Chandrasekhar1931} S. Chandrasekhar,  The density of white dwarf stars, Philos. Mag. 11 (1931) 592.

\bibitem{Chandrasekhar1935} S. Chandrasekhar,  The highly collapsed configurations of a stellar mass, Mon. Not. R. Astron. Soc. 170 (1935) 405.

\bibitem{Shapiro1983} S.L. Shapiro, S.A. Teukolsky, Black Holes, White Dwarfs and Neutron Stars: The Physics of Compact Objects : John Wiley \& Sons, New York, 1983.

\bibitem{Shah2015a} M.G. Shah, Quantum positron-acoustic waves in dense plasmas (Lap-lambert Publishing, Germany, 2015). ISBN- 978-3-659-81624-6

\bibitem{Shah2015b} M.G. Shah, M.R. Hossen, A.A. Mamun, Nonlinear propagation of positron-acoustic waves in a four component space plasma, J. Plasma Phys. 81 (2015) 905810517.

\bibitem{Eviatar 1983} A. Eviatar, R.L. Mcnutt, G.L. Siscoe, J.D. Sullivan, Heavy ions in the outer Kronian magnetosphere, J. Geophys. Res. 88 (1983) 823.

\bibitem{Richardson 1986} J.D. Richardson, Thermal ions at Saturn: Plasma parameters and implications, J. Geophys. Res. 91 (1986) 1381.

\bibitem{Ema2015} S.A. Ema, M.R. Hossen, A.A. Mamun, Nonplanar shocks and solitons in a strongly coupled adiabatic plasma: the roles of heavy ion dynamics and nonextensitivity, Cotrib. Plasma Phys. 55 (2015) 596.

\bibitem{Hossen2014a} M.R. Hossen, L. Nahar, A.A. Mamun, Roles of arbitrarily charged heavy ions and degenerate plasma pressure in cylindrical and spherical IA shock waves, Phys. Scr. 89 (2014) 105603.

\bibitem{Chatterjee2012} P. Chatterjee, D.K. Ghosh, B. Sahu, Planar and nonplanar ion acoustic shock waves with nonthermal electrons and positrons, Astrophys. Space Sci. 339 (2012) 261.

\bibitem{Mannan2012} A. Mannan, A.A. Mamun, Planar electron-acoustic solitary waves and double layers in a two-electron-temperature plasma with nonthermal ions, Astrophys. Space Sci. 340 (2012) 109.

\bibitem{El-Taibany2007} W.F. El-Taibany, M. Wadati, Nonlinear quantum dust acoustic waves in nonuniform complex quantum dusty plasma, Phys. Plasmas 14 (2007) 042302.

\bibitem{Zobaer2012} M.S. Zobaer, N. Roy, A.A. Mamun, DIA solitary and shock waves in dusty multi-ion dense plasma with arbitrary charged dust, J. Mod. Phys. 3 (2012) 755.

\bibitem{Samanta2013} U.K. Samanta, A. Saha, P. Chatterjee, Bifurcations of nonlinear ion acoustic travelling waves in the frame of a Zakharov-Kuznetsov equation in magnetized plasma with a kappa distributed electron, Phys. Plasmas 20 (2013) 052111.

\bibitem{Saha2014} A. Saha, N. Pal, P. Chatterjee, Dynamic behavior of ion acoustic waves in electron-positron-ion magnetoplasmas with superthermal electrons and positrons, Phys. Plasmas 21 (2014) 102101.

\bibitem{Shah2015m} M.G. Shah, M.R. Hossen, A.A. Mamun, Nonplanar positron-acoustic shock waves in astrophysical plasmas, Braz. J. Phys. 45 (2015) 219.

\bibitem{Shahmansouri2013} M. Shahmansouri, H. Alinejad, Effect of electron nonextensivity on oblique propagation of arbitrary ion acoustic waves in a magnetized plasma, Astrophys. Space Sci. 344 (2013) 463.

\bibitem{Wang2009} Y. Wang, Z. Zhou, Y. Lu, X. Ni, J. Shen, Y. Zhang, Relativistic magnetosonic solitary wave in magnetized multi-ion plasma, Commu. Theor. Phys. 51 (2009) 1121.

\bibitem{Masood2010} W. Masood, B. Eliasson, P.K. Shukla, Electromagnetic wave equations for relativistically degenerate quantum magnetoplasmas, Phys. Rev. E 81 (2010) 066401.

\bibitem{rasel2014a} M.R. Hossen, L. Nahar, S. Sultana, A.A. Mamun, Nonplanar ion-acoustic shock waves in degenerate plasmas with positively charged heavy ions, High Energy density Phys. 13 (2014) 13.

\bibitem{rasel2014b} M.R. Hossen, A.A. Mamun, Electrostatic solitary structures in a relativistic degenerate multispecies plasma, Braz. J. Phys. 44 (2014) 673.

\bibitem{rasel2014c} M.R. Hossen, S.A. Ema, A.A. Mamun, Nonplanar shock structures in a relativistic degenerate multi-species plasma, Commun. Theor. Phys. 62 (2014) 888.

\bibitem{rasel2015} M.R. Hossen, A.A. Mamun, Nonplanar shock excitations in a four component degenerate quantum plasma: the effects of various charge states of heavy ions, Plasma Sci. Technol. 17 (2015) 177.


\bibitem{Chakrabarti2002} N. Chakrabarti, A. Fruchtman, R. Arada, Y. Marona, Ion dynamics in a two-ion-species plasma,  Phys. Lett. A 297 (2002) 92.

\bibitem{Tskhakaya2005} D. Tskhakaya, S. Kuhn, Boundary conditions for the multi-ion magnetized plasma-wall transition, J. Nucl. Mater. $337-339$ (2005) $405-409$.

\bibitem{Haider2014} M.M. Haider, T. Ferdous, S.S. Duha, The effects of vortex like distributed electron in magnetized multi-ion dusty plasmas, Cent. Eur. J. Phys. 12 (2014) 701.


\end{thebibliography}
\end{document}